\documentclass[reprint,
superscriptaddress,
 amsmath,amssymb,
 aps,
prb,
floatfix,
twocolumn
]{revtex4-2}
\usepackage{courier}
\usepackage{graphicx}
\usepackage{dcolumn}
\usepackage{bm}
\usepackage{listings}
\usepackage{xcolor}
\usepackage{color}
\usepackage{array}
\lstdefinestyle{stylepython}{
    language=Python,
    basicstyle=\ttfamily,
    commentstyle=\color{green},
    keywordstyle=\color{blue},
    stringstyle=\color{olive},
    numberstyle=\tiny,
    numbers=left,
    stepnumber=1,
    numbersep=5pt
}

\begin{document}


\title{Analysis and fabrication of antireflective coating for photovoltaics based on a photonic-crystal concept and generated by evolutionary optimization}


\author{Pauline Bennet}
\affiliation{Universit\'e Clermont Auvergne, CNRS, SIGMA Clermont, Institut Pascal,F-63000 Clermont-Ferrand, France}
\author{Perrine Juillet}
\affiliation{Universit\'e Clermont Auvergne, CNRS, SIGMA Clermont, ICCF, F-63000 CLERMONT-FERRAND, FRANCE}
\author{Sara Ibrahim}
\affiliation{Universit\'e Clermont Auvergne, CNRS, SIGMA Clermont, Institut Pascal,F-63000 Clermont-Ferrand, France}
\author{Vincent Berthier}
\affiliation{TAO, Inria, LRI, Universit\'e Paris Sud CNRS UMR 6823, Orsay Cedex, France}
\author{Mamadou Aliou Barry}
\author{François R\'everet}
\affiliation{Universit\'e Clermont Auvergne, CNRS, SIGMA Clermont, Institut Pascal,F-63000 Clermont-Ferrand, France}
\author{Ang\'elique Bousquet}
\affiliation{Universit\'e Clermont Auvergne, CNRS, SIGMA Clermont, ICCF, F-63000 CLERMONT-FERRAND, FRANCE}
\author{Olivier Teytaud}
\affiliation{Facebook AI Research, 6 rue Menars, 75000 Paris, France}
\author{Emmanuel Centeno}
\author{Antoine Moreau}%
\affiliation{Universit\'e Clermont Auvergne, CNRS, SIGMA Clermont, Institut Pascal,F-63000 Clermont-Ferrand, France}


\date{\today}

\begin{abstract}
We optimize multilayered anti-reflective coatings for photovoltaic devices, using modern evolutionary algorithms. We apply a rigorous methodology to show that a given structure, which is particularly regular ({\em i.e. essentially periodic}), emerges spontaneously and systematically for a very broad range of conditions. The very regularity of the structure allows for a thorough physical analysis of how the design operates. This allows to understand that the central part is a photonic crystal utilized as a buffer for light, and that the external layers have the purpose of reducing the impedance mismatch between the outer media and the Bloch mode supported by the photonic crystal. This shows how optimization can suggest new design rules and be considered as a source of inspiration. Finally, we fabricate these structures with easily deployable techniques.
\end{abstract}

\maketitle

Numerical optimization of photonic devices is increasingly relied on in the literature to provide us with solutions to technological needs, like highly miniaturized multiplexers\cite{piggott2015inverse}. This constitutes a path towards efficient solutions when no clear design rules are available. Such an approach clearly dates back to two decades ago, when particularly efficient design tools for multilayered structures have been developed\cite{tik1994,tik1996}. More recently, advanced optimization methods for photonic devices have been proposed and successfully used to generate numerous photonic structures\cite{molesky2018inverse}.

However, no thorough physical analysis of the response of the structure is generally done, for the simple reason that the structures are way too complex to be truly understood. The algorithms present for instance a well known tendency to generate extremely small features which make them more difficult to fabricate and commercialize\cite{molesky2018inverse}. The results of these optimizations do not suggest any new design rule, so that numerical optimization is not yet considered as a true source of inspiration. This is all the more frustrating that naturally occurring photonic structures do inspire original designs\cite{potyrailo2007morpho}, which is made possible by the fact that these photonic structures are almost systematically regular in some way or another\cite{kinoshita2008physics}.

In this paper, we will call structures {\em regular} when they may be periodic, have periodic parts or present a gradual change. In that sense, regular structures are obviously structures which can be physically analyzed and understood. We underline however that the link between regularity  and our comprehension is even more profound: when regularity can be identified in the most disordered structures or problems\cite{moreau2006optimal,yu2020machine} then it can be interpreted physically. The spontaneous emergence of regularity in optimization should thus really be sought after and never considered lightly -- this may be a sign that design rules can be derived from optimization results. We underline that in other domains, the same methods which are used in photonics actually yield regular solutions and are thus considered as a source of inspiration\cite{liu2018current}. 

In photonics however, the emergence of regularity is much less common, which can be linked to the numerous resonances even the most simple structures may present : each resonance represents potentially a local minimum for a cost function, and thus an increase in the complexity of the problem for an optimization algorithm. It is then difficult to determine whether regular structures do not emerge spontaneously because of the method used to perform the optimization or because regular structures are simply not optimal. A pioneer work however has suggested more than a decade ago that choosing the right optimization algorithm is important, by showing that 
evolutionary algorithms, which are versatile and global optimization methods, are able to generate regular photonic structures spontaneously. More recently, it has been shown that such algorithms are indeed perfectly able to retrieve naturally occurring photonic structures in the most simple yet emblematic cases\cite{barry2018evolutionary}.

When applied to the problem of designing a multilayered anti-reflective coating for a photovoltaic device, these evolutionary algorithms yield a particularly elegant and regular solution based on a photonic crystal with obviously adapted external layers. 
The emergence of a partially periodic solution may seem at first surprising. Photonic crystals may present a forbidden band, and thus behave as a frequency selective mirror -- that is why in all the other optimization problems they emerge when the reflection, not the transmission, is maximized\cite{barry2018evolutionary}. The optimization results thus suggest that photonic crystals may be utilized when they are transparent as efficient anti-reflective coatings. 


Here we show that this structure emerges as the result of the optimization for a broad range of conditions and in a very systematic way. Then, we conduct a thorough physical study of this anti-reflecting coating based on a photonic crystal, which is made possible by the very regularity of the solution proposed by the optimization algorithms. 

In a first part, we study in which conditions the structure may emerge, and show that they are broader than initially thought, explaining why it is consistently proposed as a solution by the optimization. A simple physical analysis then allows us to discuss to which point the structure is actually optimal. In a second part, we provide a thorough physical analysis of the optical response of the structure. Our analysis is based on a Fabry-Perot model for the Bloch mode propagating inside the crystal, and we provide tools to compute the reflection coefficient of the Bloch mode inside the structure. We show that the photonic crystal is used, outside its forbidden band, as a buffer allowing light coming from free space to be transferred to silicon in a very efficient way. Finally, we have checked that such a structure could actually be realized using techniques which would be easy to deploy in the industry.

\section{Emergence of the design}

The structure we study in the present work has emerged as a solution consistently produced by global optimization algorithms when asked to find the best possible Anti-Reflective Coating (ARC) in normal incidence as a multilayer with two alternating refractive index (1.4 and 1.7 which are typical of the silicon oxynitride we use in the last part of the paper to fabricate the ARC\cite{farhaoui2016reactive} or of biological materials, typically chitin\cite{parker2001}) on top of amorphous hydrogenated silicon layer. 

The quantity to be maximized here was the short-circuit current in the 375-750 nm range, assuming a quantum yield equal to 1, as described in \cite{defrance16}. Among the different algorithms which were tested, Differential Evolution (DE) proved to be the most efficient algorithm for this kind of problem, generating solutions faster and more reliably than any other global algorithm\cite{barry2018evolutionary}.

\begin{figure}
    \centering
    \includegraphics[width=\linewidth]{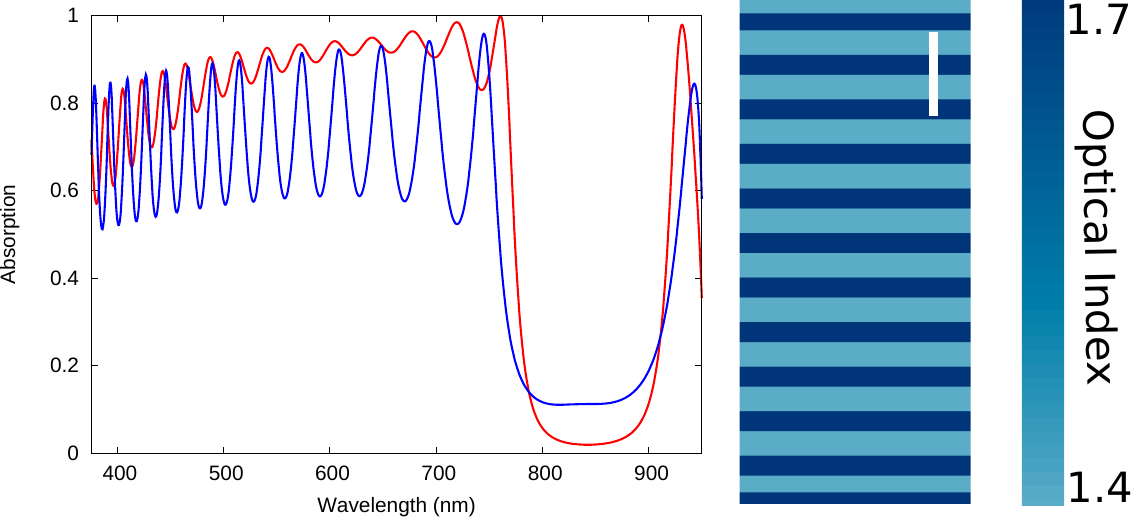}
    \caption{Absorption inside the silicon layer for the optimized design (red) and of the pure photonic crystal (blue). The optimized design minimizes the ripple - the amplitude oscillations of the absorption - in the visible range. The white line represents 500 nm. }
    \label{fig:absorption}
\end{figure}

\begin{figure}
    \centering
    \includegraphics[width=\linewidth]{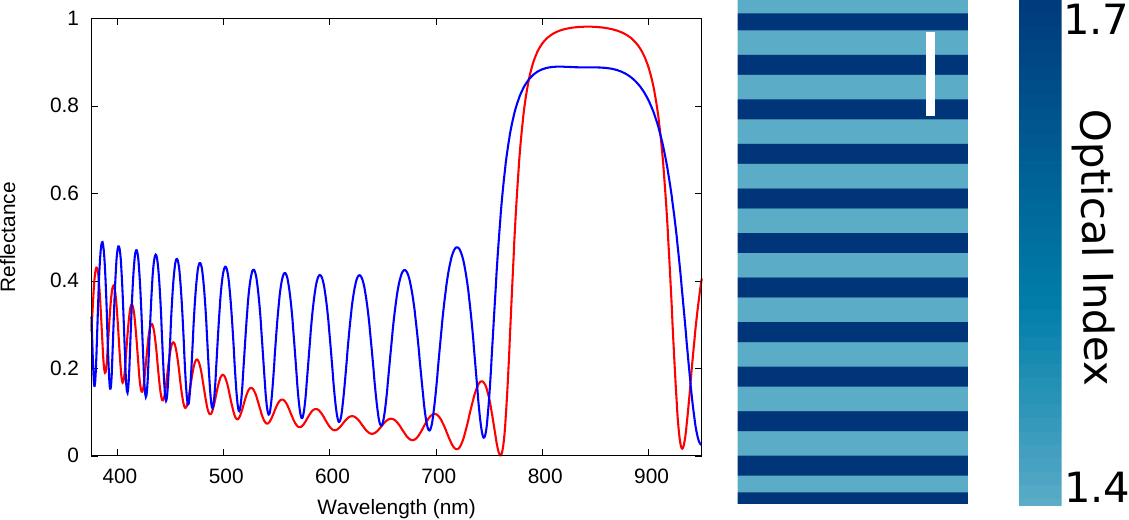}
    \caption{Reflectance of the optimized design (red) and of the pure photonic crystal (blue). The optimized design represented minimizes the ripple - the amplitude oscillations of the reflectance - in the visible range. The white line represents 500 nm.}
    \label{fig:reflection} 
\end{figure}

The structure which will serve as a reference in the following, and that we call the {\em simplified ARC} is constituted of a Bragg mirror with a forbidden band center at 835 nm, with two adapted layers on top of the structure (in contact with air) of 79 and 102 nm; and at the bottom (close to silicon) of 100 and 70 nm. The structure as well as its absorption spectrum are shown on Fig. \ref{fig:absorption}, compared to the absorption spectrum of the photonic crystal alone. The role of the additional layers can be easily understood: they reduce the oscillations which can be seen in the spectrum and thus maximize the absorption in the visible range where the silicon is absorbent. The overall behavior of the structure is even better understood when looking at the reflectance of the structure, because the reduction of the oscillations and the high reflectance due to the forbidden band clearly appear on the spectrum Fig.\ref{fig:reflection}. This structure evokes Bragg mirrors equipped with ripple reduction layers, a classical technique used in optical filter synthesis for Bragg mirrors deposited on a glass substrate\cite{macleod2001thin,macleod1,macleod2,macleod3}. Such a technique is however not used to generate anti-reflective coatings, as this is generally not the goal which is pursued.

In this part, we will discuss the optimally of the ARC structure, keeping in mind that it is generally impossible to prove that a solution provided by an optimization algorithm is actually optimal. Even worse, one should assume that {\em an optimization almost never produces an optimal design}. It is possible to build some confidence that a solution is close to being optimal when for instance (i) it is consistently obtained as a solution by the optimization algorithms (ii) a simple physical analysis allows to understand how the design operates. Here, we will couple a statistical analysis of the emergence of our solution to our physical understanding of this solution. We show that this kind of structure is not optimal, by providing slightly more efficient ARC for the largest index contrasts. However, as we show here, the fact the same structure emerges consistently reinforces our confidence that it is close to being optimal.

\subsection{Methodology of the optimization}

The cost function is the function which the algorithm will try to minimize. In our case, for a given multilayer structure characterized by the refractive index and thickness of each layer, we compute the short circuit current assuming an AM1.5 standard solar spectrum $j_{sc}$ using the formula
\begin{equation}\label{eq:jsc}
j_{sc} = \int A(\lambda) \frac{d I}{d\lambda}.
 \frac{\mbox{e}\lambda}{\mbox{h} \mbox{c}}\,\mbox{d}\lambda
\end{equation}
where $A$ is the absorbance of the active layer, e, h and c are respectively the elementary charge, the Planck constant and the speed of light in vacuum and the spectral density of the illumination is given the solar spectrum\cite{santbergen2010am1}.  We divide $j_{sc}$ by the maximal theoretical short-circuit over the 375-750 nm range, $j_{max}$. Since we want to maximize this quantity, the cost function $f$ is defined by $$f = 1-\frac{j_{sc}}{j_{max}}$$
so that it is comprised between 0 and 1. Said otherwise, when the cost function is decreased by 0.1, this corresponds to an increase of the short-circuit current equal to 10\% of the maximum achievable current.

We expect the problem we study to be modular, meaning that different regions of the structure will play almost independent roles, which is often the case for multilayered structures. In order to chose the right algorithm as well as the right variant and parameters for our problem, we have compared different algorithms and their variants on the most modular problem we have identified: the case of the high reflectance filter on a broad spectral range, leading to a chirped design\cite{chirped97}. On this problem, the evolutionary algorithm called Differential Evolution (DE) is the most efficient\cite{barry2018evolutionary}, but several variants of DE exist. After comparing the most common variants (\texttt{DE/rand/1}, \texttt{DE/best/1}, \texttt{DE/randToBest/1}, \texttt{DE/currToBest/1}, \texttt{DE/rand/2}, \texttt{DE/best/2}, identified according to the usual classification \cite{de}) we have selected the \texttt{DE/randToBest/1} variant. Finally, we have tuned the parameters of the algorithm so that it is finds the solution are quickly and as reliably as possible for the chirped dielectric mirror. We underline that this testbed is freely available in the Nevergrad library\cite{nevergrad,bennet:hal-02613161}. 

We thus consider a population of 30 individuals chosen randomly. Each individual is a potential structure solution and is defined by a vector, corresponding to a location in what we call here the phase space. To generate a new generation, each individual is compared to a new individual. The parameters of the new individual come either directly from the current individual to be improved with a probability $cr=0.2$ or from a mutant generated using the formula
\begin{equation}
    y=c+F_1\,(a-b)+F_2\,(best-c)\label{eq:de}
\end{equation}
where $y$ is the mutant, $a$, $b$ and $c$ are the vectors which correspond to individuals chosen randomly in the population, while $best$ is the vector corresponding to the individual with the lower cost function. We have taken $F_1=0.9$ and $F_2=0.8$. The cost function associated to the new individual is computed, and it replaces the current individual in the population only if it has a lower cost function.

We have put boundaries on all the parameters. Here we force the thicknesses of each layer to be comprised between 30 and 250 nm. When a new individual is generated using mutation, there is no guarantee that it will satisfy these conditions. When this is not the case for a parameter, it is simply replaced by the parameter of the current individual. This allows to explore correctly all the authorized phase space, without neglecting the edges -- because the best solution is frequently found near these limits, especially when the refractive index vary\cite{barry2018evolutionary}. We underline that in some of our optimizations, we left the refractive indexes of each layer vary. In those cases we performed a structured optimization by considering that the refractive index and the thickness of a layer constitute a single parameter: when a new individual is generated, either the two parameters are taken from the parent or they are computed using formula \eqref{eq:de} with the same individuals. In general this drastically improves the efficiency of the algorithm\cite{rapin2020open}, underlining once more what a physical understanding can bring to the optimization even {\em a priori}.

\begin{figure}
    \centering 
    \caption{Best cost functions obtained with DE for the optimization of the ARC on an infinite amorphous silicon substrate for a structure with 12 layers (red data) and 46 layers (blue data) sorted from the most efficient toward the less efficient result. The best result obtained for 12 layers is consistently obtained contrarily to the 46-layers solution.}
      \includegraphics[width=1\linewidth]{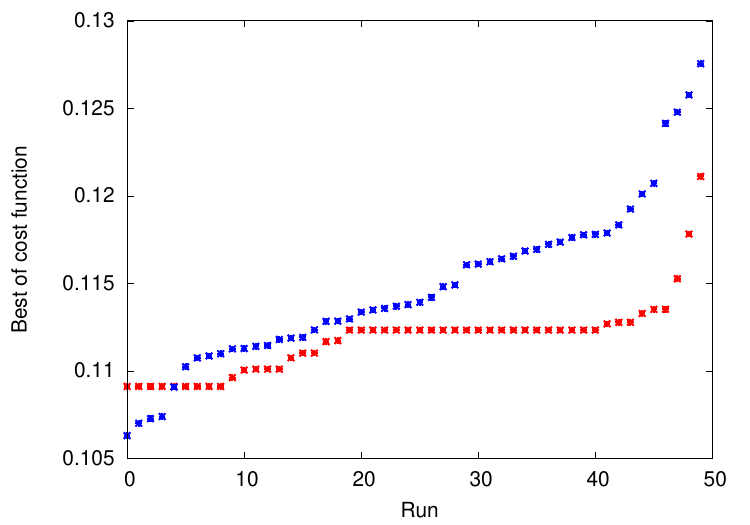} 
    \label{fig:comp_fiab_reference}
\end{figure}

In order to estimate whether a solution is produced systematically or not, we run 50 times DE on each problem (with a given number of layers), making sure that each run has converged by allocating a large enough budget\cite{supp}. Then we plot all the different values of the cost function for the best individual obtained for each optimization, sorting them starting with the lowest as shown Fig. \ref{fig:comp_fiab_reference}. 

This kind of representation allows to estimate whether the solution was produced by pure luck or not. A plateau for the lowest value indicates that the solution is often found. A plateau at a different value indicates that there is another solution with a large attraction basin in the phase space which is often found by the algorithm (these two types of plateau can be seen Fig. \ref{fig:comp_fiab_reference}).

Generally, for a multilayered optical filter, the higher the number of layers, the more efficient it is, and thus the lower the cost function \cite{macleod2001thin,tik1996,bennet:hal-02613161,barry2018evolutionary}. This unique property allows to better estimate when the algorithms struggle to find a solution by plotting the lowest cost function obtained for 50 run versus the number of layers of the structure. In general, this curves shows a smooth decline as the number of layers grows and then becomes chaotic -- signaling that the results are no longer reliable because the problem is too complex for the algorithm. 

\subsection{Results}

\begin{figure}
    \centering
    \includegraphics[width=\linewidth]{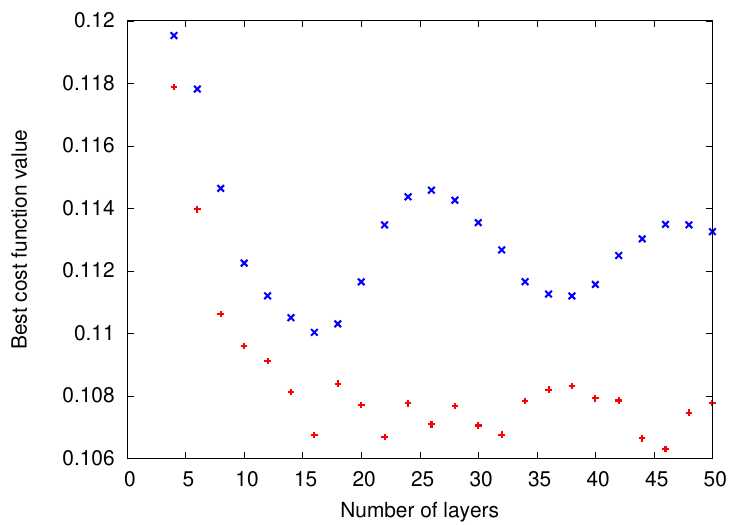} 
    \caption{Best value of the cost function obtained among 50 optimizations with DE algorithm in function of the number of layers in the structure, for the standard conditions with a silicon substrate (red curve). 
    The blue curve represents the value of the cost function for the {\em simplified ARC} that is constantly suggested by the algorithms: a central photonic crystal with adapted layers.} 
    \label{fig:mod_ref_Sia}
\end{figure}

First we have run optimizations as described above, for an even number of layers ranging from 4 to 50, the structure beginning with the lower index (1.4) and ending with the higher (1.7). The structure is placed on top of an infinite hydrogenated amorphous silicon substrate. 
We have plotted Fig. \ref{fig:mod_ref_Sia} the best cost function obtained as a function of the number of layers. On the same figure, we have plotted the cost function obtained for our {\em simplified ARC}, described above.


Changing the thickness of the photonic crystal inside the ARC amounts to shifting the oscillations which can be seen on its spectrum as shown Fig.\ref{fig:reflection}. As a result, the cost function of the {\em simplified ARC} presents quasi-periodic oscillations when the number of layers increases. The values of the cost function for the {\em simplified ARC} are just slightly larger than for the fully optimized structure. We consider this is a sign that small changes are brought by the algorithms to adapt even more the ARC, but that they will bring minor improvements only. The {\em simplified ARC} thus captures the main characteristics of the optimized structures. Furthermore, Fig. \ref{fig:mod_ref_Sia} indicates that even for 50 layers the optimization is reliable, since it always provides solutions that are better than the {\em simplified ARC} despite the increase in complexity of the problem.

Setting the refractive index of the layers, how they alternate, an even number of layers are arbitrary choices which may have introduced bias and may be responsible for the emergence of the structure. We have thus run optimizations for different cases\cite{supp} and almost each time, the optimized structure is a photonic crystal with the most of the time two adapted layers at each sides. 

\begin{figure}
    \centering
    \includegraphics[width=\linewidth]{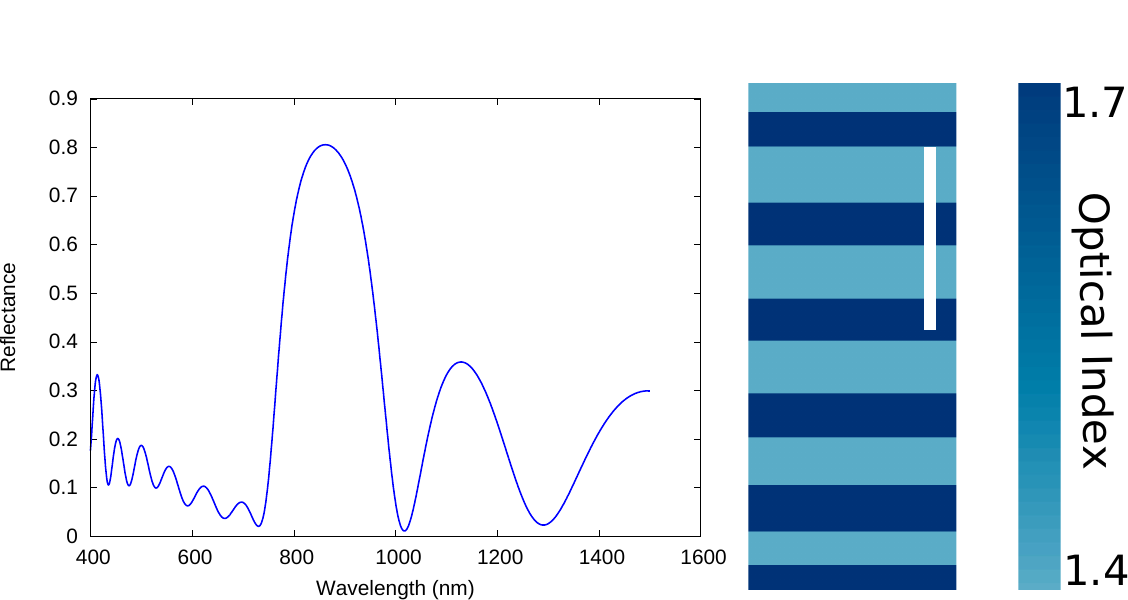}
        \caption{Reflectance spectrum and representation of the best structure optimized with 12 layers. The white line represents 500 nm.}
    \label{fig:spectrum_structure_12_ref_Sia}
\end{figure}

We represent the cost function values of the best structures obtained in the standard conditions as a function of the number of layers in the structure in red in Fig.\ref{fig:mod_ref_Sia}. We note that the structure with 12 layers is particularly efficient. Fig. \ref{fig:spectrum_structure_12_ref_Sia} shows the structure and the corresponding reflectance spectrum. Fig. \ref{fig:comp_fiab_reference} shows that this 12 layers structure is obtained very consistently, since a plateau can be seen for the lowest values of the cost function.

\subsection{Discussion}

Our results clearly indicate that a Bragg mirror with adapted outer layers is an obvious strategy to reduce the reflection for a photovoltaic device. The design emerges systematically whatever the conditions. Even when the algorithm cannot be considered to produce a satisfactory solution ({\em i.e.} as efficient as the {\em simplified ARC}), the results point in that direction, obviously resembling the reference and presenting the same level of efficiency but never clearly outperforming it.

\begin{figure}
\includegraphics[width=1\linewidth]{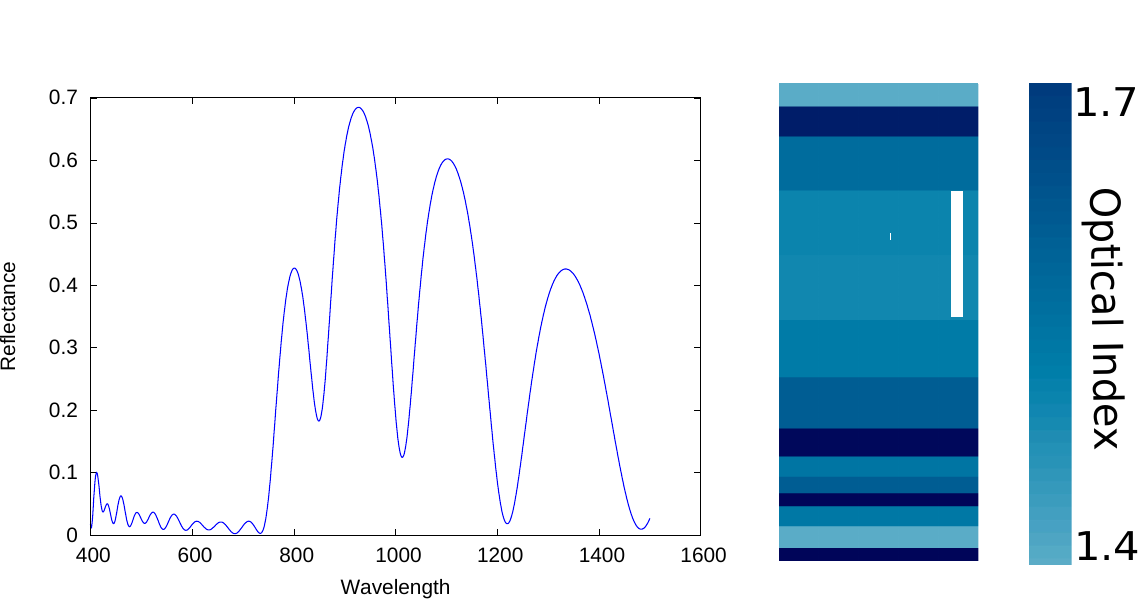}
    \caption{Reflectance spectrum and representation of the structure with 14 layers, when the refractive index can vary on a large contrast. The white line represents 500 nm.}
    \label{fig:spectrum_structure_14_GC_nvar}
\end{figure}

For a photovoltaic problem, extending over most of the visible spectrum and with a relatively bumpy AM1.5 illumination, a much more disordered structure could have been expected. In similar situations, engineered disorder has actually proven to be a very fruitful approach\cite{yu2020engineered}. Even seemingly ordered structures which occur in nature, like that of the Morpho butterfly\cite{barry2018evolutionary} do contain some kind of correlated disorder\cite{yu2020engineered}. The structures we obtain do also, to better adapt to the AM1.5 spectrum obviously, but this remains limited -- so that the ordered structure can always be distinguished. However, different structures may well emerge if the way the efficiency of the structure is assessed (and thus the cost function) changes. Figure \ref{fig:spectrum_structure_14_GC_nvar}
shows another ordered design, which emerges when the refractive index range is extended and the refractive index is let to vary. This structure can be deemed regular, with a striking symmetrical pattern at its core. However, the reasons why it emerges remain unclear: one would expect such a smooth refractive index profile to lead to some kind of localization of light inside the structure, a seldom desirable feature in this context. This shows anyway that while the range of situations where a periodic structure emerges is relatively broad, such a structure is not systematically the best solution.

Moreover, simulations show that when the incidence angle is modified, the photonic bandgap of the crystal is shifted towards shorter wavelengths, which is detrimental to the efficiency. A new cost function including a wide angular range may then lead to completely different structures, eventually disordered. The design we present may thus be optimal only in normal incidence.

We underline that the solutions to ARC problems in the literature, generally obtained using specific design algorithms\cite{tik1994,tik1996}, are generally complex, but sometimes quite regular too\cite{tikhonravov2008estimation} -- so that the emergence of a periodic design is not new. However, the designs are still difficult to understand compared to the design we study here and which is definitely original.

\begin{figure}
    \centering
    \includegraphics[width=\linewidth]{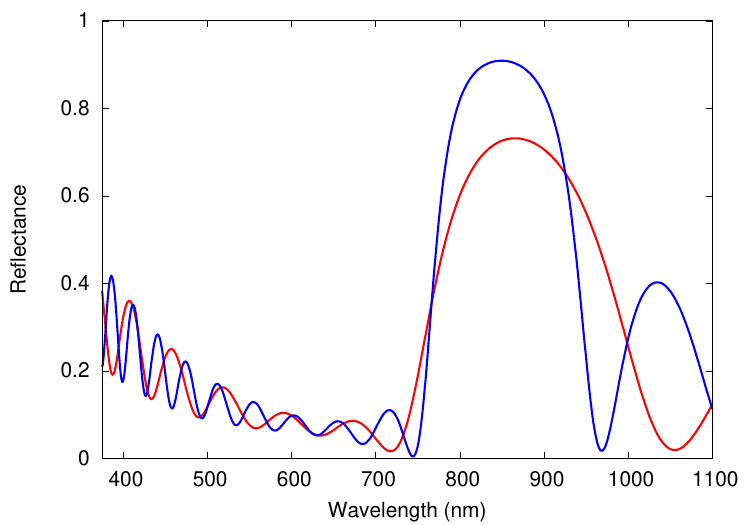}
    \caption{Reflectance of the {\em simplified ARC} with 6 layers (red) and 12 layers (blue) in the central photonic bandgap. The number of layers does not impact the ripple amplitude.}
    \label{fig:ref_6_12_layers}
\end{figure}

The strategy suggested here by the optimization is very similar to what is done for Bragg mirrors in order to reduce the so called ripples -- the oscillations in the spectrum outside of the bandgap\cite{macleod1,macleod2,macleod3,Thelen:66, Thelen:71}. Such oscillations are commonly thought as produced by some kind of reflection inside the structure due to the "mismatch between the structure and the surrounding media"\cite{macleod2}. The amplitude of the ripples is actually not modified whatever the number of layers of the photonic crystal, as shown Fig. \ref{fig:ref_6_12_layers}. We have thus followed this idea by trying to retrieve the reflection coefficient of the Bloch mode inside the structure and to see how it is modified by the adapted layers.

\section{Physical analysis through a cavity model for the Bloch mode}

Inspired by the results provided by the optimization, we study in this part our {\em simplified ARC} as described above.

Even though a Bragg mirror is actually considered by most authors as a slab in which a Bloch mode propagates and is reflected by both edges, there has been little work supporting such an idea because of the lack of any theoretical method to define or compute the reflection or transmission coefficients of the Bloch mode.

Here we show that such coefficients can be computed by using a cavity model to fit the actual data, allowing a thorough physical analysis of the whole structure.

\subsection{Fabry-Perot cavity model}

\begin{figure}
    \centering
    \includegraphics[trim=5cm 3cm 5cm 0cm, clip,scale=0.4]{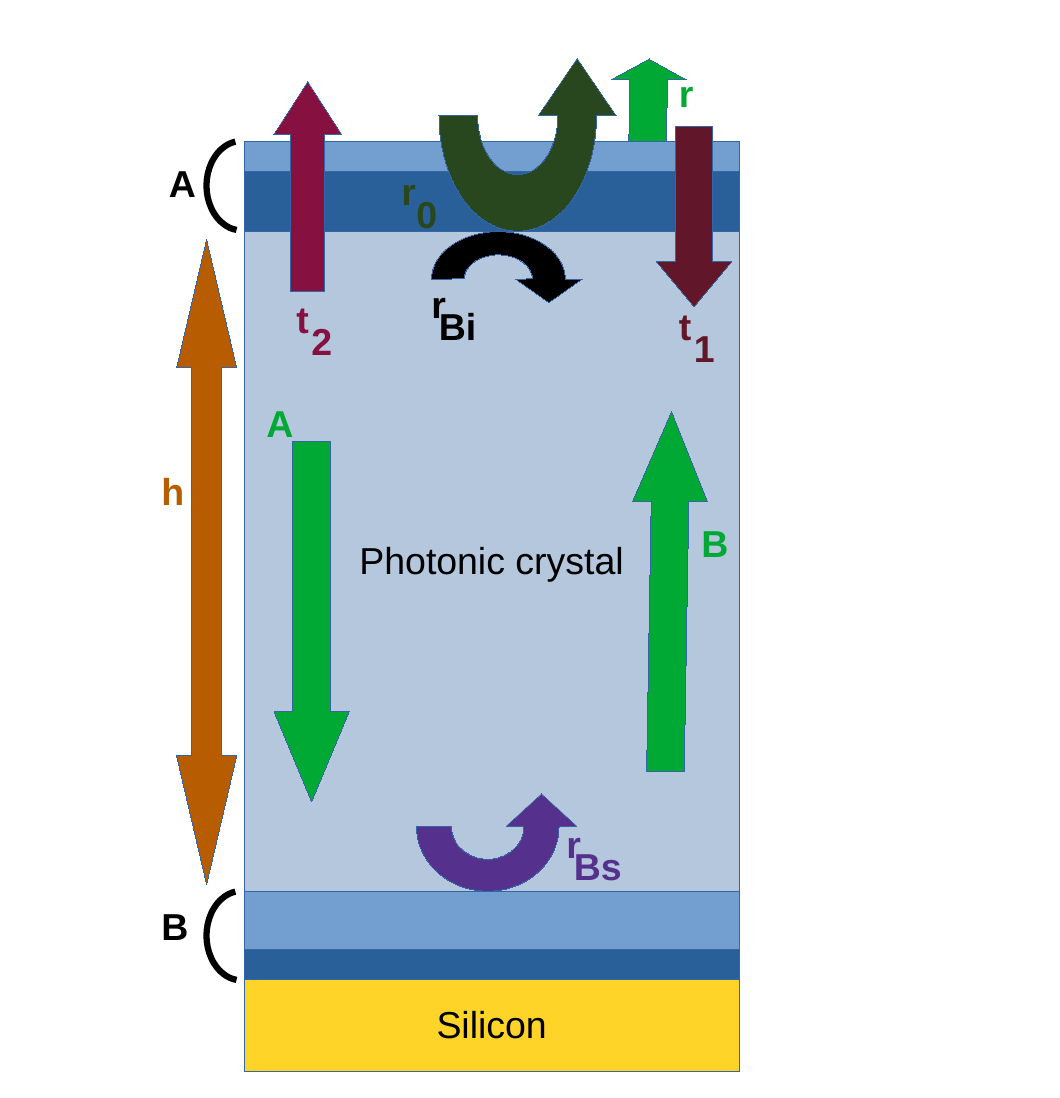}
    \caption{Cavity model for the reference design. A: air-adapted layers. B: silicon-adapted layers.
    $r_{Bi}$: reflection coefficient of the Bloch mode on the upper adapted layers. $r_{Bs}$: reflection coefficient on the lower (silicon side) adapted layers. $t_1$: transmission coefficient from the incidence wave to the Bloch mode. $t_2$: transmission coefficient from the Bloch mode to the plane wave. $r_0$: reflection coefficient of the incoming plane wave on the first layer of the photonic crystal.}
    \label{fig:simple_cavity}
\end{figure}

Inside an infinite Bragg mirror, light propagates as in an homogeneous medium under the form of a Bloch mode, which has the profile of a plane wave. Each Bloch mode is characterized by a wavevector $k_x$ along the direction of the interfaces and a wavevector perpendicularly to the interfaces noted $\gamma$ here. In the following, the light is assumed to illuminate the structure in normal incidence, so that $k_x=0$ and the dispersion relation giving $\gamma$ reduces to
\begin{equation}
    \cos(\gamma d)=\cos \phi_1 \cos \phi_2 -\frac{\frac{n_1}{n_2} + \frac{n_2}{n_1}}{2} \sin \phi_1 \sin \phi_2,
    \label{eq:gamma}
\end{equation}
with $d=d_1+d_2$ the period of the photonic crystal, $\gamma_i$ the wavevector inside medium $i$ of refractive index $n_i$ and thickness $d_i$, and $\phi_i=\gamma_i d_i$. 

In the bandgap of the Bragg mirror, $\gamma$ is purely imaginary, which means that the Bloch mode will not propagate inside the structure. In the spectral range we consider, the wavevector will be assumed to be real and the Bloch mode propagative.

A Bragg mirror with a finite number of periods, with or without adapted outer layers will be considered here as a cavity for the Bloch mode defined above, with a wavevector $\gamma$ given by the dispersion relation for the infinite Bragg mirror. 

In the framework of such a model, the reflection coefficient $r$ of the whole structure with a Bragg mirror with a thickness $h$ is given by a cavity model 
\begin{align}
    & r=r_0+t_2 B, \\
    & A=t_1+Br_{Bi}, \\
    & B=Ar_{Bs}e^{2i\gamma h},
\end{align}
with $A$ and $B$ the upward and downward amplitudes of the Bloch mode, $r_0$ the reflection of the incident plane wave on the first layer of the photonic crystal, $t_1$ and $t_2$ the transmission coefficients between free space and the Bloch mode, $r_{Bi}$ (resp. $r_{Bs}$) is the reflection coefficient of the Bloch mode on the outer layers (resp. the substrate). The whole model with the different coefficients is summarized Fig. \ref{fig:simple_cavity}.

Those formula leads to Fabry-Perot formula for the reflection coefficient of the whole structure
\begin{equation}
    r(\lambda, h)=r_0(\lambda)+\frac{t_1(\lambda) t_2(\lambda) r_{Bs}(\lambda) e^{2i\gamma(\lambda) h}}{1-r_{Bi}(\lambda) r_{Bs}(\lambda) e^{2i\gamma(\lambda) h}}.
        \label{eq:model}
\end{equation}

\subsection{Fit with a Fabry-Perot formula} \label{Section:Fit}

\begin{figure} 
    \centering
    \includegraphics[width=\linewidth]{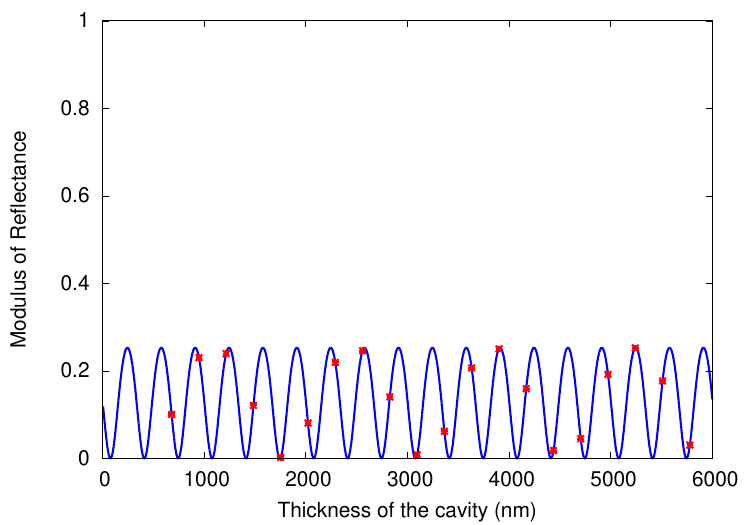}
    \caption{Reflectance obtained at 692 nm for different thicknesses of a symmetric photonic crystal terminated by the lower index layers and with air on both sides (red crosses) and fit (blue curve) by a Fabry-Perot formula.} 
    \label{fig:fit} 
\end{figure}

Retrieving the internal transmission and reflection coefficients of the model would be a relatively easy task if only it were possible to make the thickness of the structure take arbitrary values for a given frequency. However, the core of the structure has a thickness which is always a multiple of the period of the crystal $d$. The problem is illustrated Fig. \ref{fig:fit} where the red point indicate the available data, as well as a reliable fit with  a periodicity imposed by the Bloch mode wavevector. Retrieving the internal coefficients is thus non trivial and requires a purely numerical approach.

To summarize, we fit the data with a cavity model. However, given the structure of the formula, it is not possible with a single fit to determine all the internal coefficients. The model for the reflectance which we actually use is given by 
\begin{equation}
    F_j=\left |  c_0 + \frac{c_1\, e^{2i\gamma h}}{1-c_2\, e^{2i\gamma j d}} \right |^2,
\end{equation}
where $c_0$, $c_1$ and $c_2$ are the coefficients which can actually be retrieved, $\gamma$ is given by the dispersion relation (\eqref{eq:gamma}) and $j$ is the number of periods of the photonic crystal considered. In order to perform the fit, a rather complicated optimization process, we use DE again, and the cost function, defined for a given frequency by
\begin{equation}
    f=\sum_{j=2}^{50} \left| r_j r_j^* - F_j\right|^2,
\end{equation}
where $r$ is the reflection coefficient of the whole structure, computed using an impedance formalism\cite{macleod2001thin} to reduce the computational cost. A single optimization run is usually enough to determine satisfactory $c_i$ coefficients, as shown Fig. \ref{fig:fit}. It is striking how the periodicity of the model, imposed by the wavevector $\gamma$ calculated independently of the fit, is different from the apparent periodicity of the data. This illustrate why the fitting process is difficult.

\begin{figure}
    \centering
    \includegraphics[width=0.8 \linewidth]{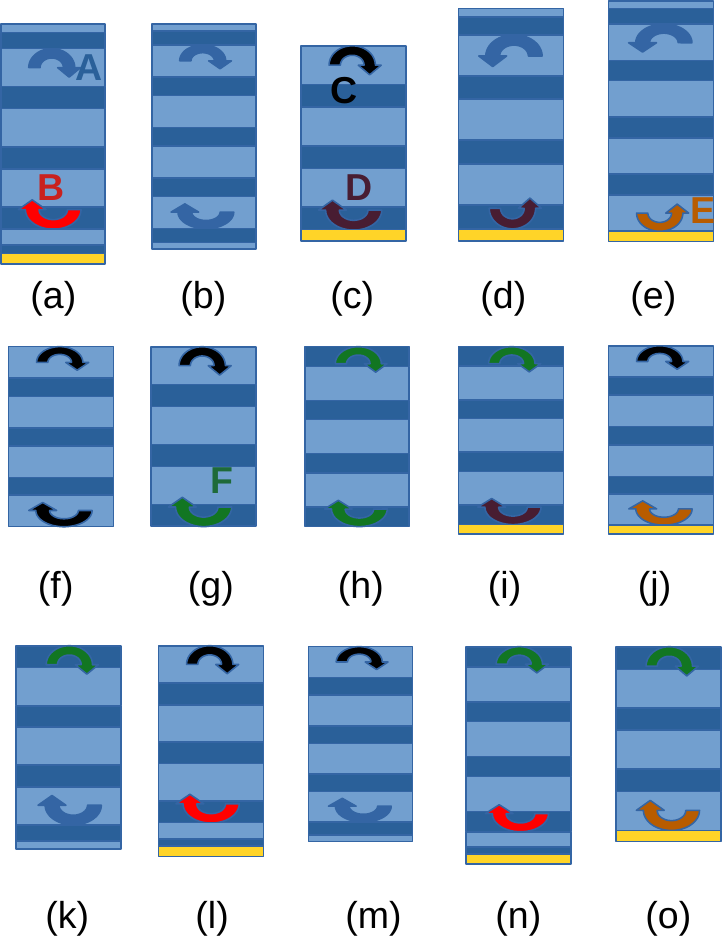}
    \caption{Representation of all the designs studied. Each curved arrow represents one Bloch mode reflection coefficient from photonic crystal toward an external media. The design of interest is this one represented on (a). The other designs allow the study of the behavior of the Bloch mode in function of the external media and to understand why the (a) design is the solution of the photovoltaic design optimization. A yellow layer represents a silicon substrate.}
    \label{fig:all_structures}
\end{figure}

We can find the $c_2$ coefficient by using the fitting process.
However, it provides an immediate access to the internal reflection coefficients only in the case of a symmetric structure. In general, $c_2$ is the product $r_{Bi} r_{Bs}$ and thus the determination of both coefficients is not straightforward. We prove\cite{supp} that determining these coefficients is possible and constitutes a coherent approach, provided the fitting process is applied to a variety of symmetric and asymmetric structures (see Fig.\ref{fig:all_structures}).

We have check for all the structures shown Fig. \ref{fig:all_structures} that our approach produces perfectly coherent results, allowing us to access the reflection coefficient of the Bloch mode whatever the termination for the Bragg mirror considered (abrupt or adapted, with silicon or with air). 

\subsection{Discussion}

Now that we have access to the reflection coefficient of the Bloch mode, we can compare the response of the structure when the Bragg mirror is terminated abruptly and when the adapted layers are added.

\begin{figure}
    \centering
    \includegraphics[scale=0.69]{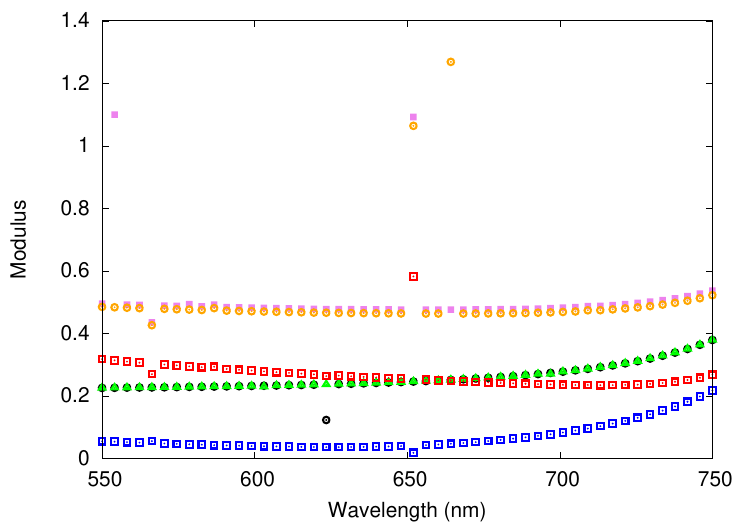}
    \caption{Coefficient reflection for the Bloch mode depending for the simple Bragg mirror (i) beginning with a high index layer (green triangles), (ii) beginning with a low index layer (black circles), (iii) beginning with adapted layers on the air side (blue squares), (iv) terminating with adapted layers on the silicon side (red squares), (v) terminating with a high index layer (pink squares), or (vi) terminating with a low index layer (orange circles). Clearly the adapted layers reduce the reflection for the Bloch mode. }
    \label{fig:rb_impact}
\end{figure}

Fig. \ref{fig:rb_impact} shows the comparison between the internal reflection coefficient with abrupt or adapted termination, whether on the side of air or of silicon. In both cases the role of the adapted layers appears very clearly: the modulus of the reflection coefficient is significantly reduced. This is particularly striking for the layers between the Bragg mirror and air. This strengthen the idea of that the adapted layers allow to reduce the mismatch between air and the Bragg mirror, reducing the internal reflection coefficient and finally the ripples.

The Bragg mirror can finally be considered as a buffer in the structure generated by the algorithms. It is difficult to reduce the impedance mismatch between air and silicon, because this mismatch is really large. Anti-reflective coatings with gradually increasing refractive index can be very efficient, but such structures require a wide range of available refractive index, typically ranging from 1 to the at least 3 (close to the refractive index of silicon)\cite{centeno2019ultra}.
The Bragg mirror with adapted terminations emerges because, with a limited range of refractive index, a better solution is to minimize the mismatch between free space and the photonic crystal, and then between the photonic crystal and silicon. The role of the Bragg mirror is thus clearly to support the Bloch mode, characterized by an effective index, and to serve as a buffer. That is the reason why, when the number of layers is increased, the algorithms tend simply to add periods to the Bragg mirror.

\begin{figure}
    \centering
    \includegraphics[scale=0.69]{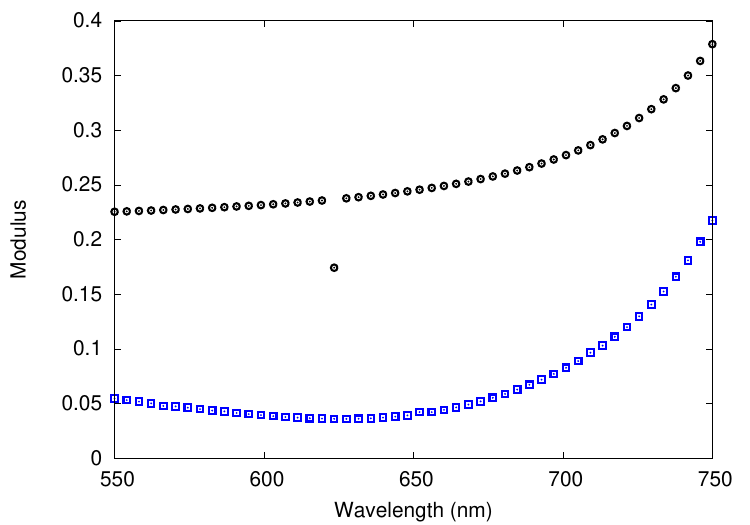}
    \caption{Impact of external layers on $r_0$. The reflection coefficient $r_0$ is diminished when adapted layers are present (blue curve) compared to when there are none (black curve).}
    \label{fig:r0_impact}
\end{figure}

This interpretation is reinforced when the $r_0$ coefficient, the reflection coefficient of the incoming light on structure with an infinite Bragg mirror layer, which can be deduced from our optimizations, is considered.
If we compare the $r_0$ coefficient obtained for structures without adapted layers, we see a reduction of $r_0$ by a factor of 4 to 5 typically when adapted layers are added (see Fig. \ref{fig:r0_impact}), which is completely coherent with the idea of a reduced impedance mismatch between air and the photonic crystal.

\section{Fabrication and characterization of the anti-reflective coating}

Finally, we have checked that the designs suggested by optimization could be easily fabricated while relying on a technique which could potentially be deployed on currently
existing production lines. We have used a reactive sputtering, a process which is known to be easily scalable, to fabricate non-stoechiometric silicon oxynitride $Si_xO_yN_z$ layers whose refractive index range typically from 1.4 to 1.7. Higher index can be reached for a higher content in silicon with this technique\cite{farhaoui2016reactive}, however in that case the layers become absorbent. A crystalline silicon target is placed at 9.5 cm in front of a rotating substrate holder 
and sputtered using a radiofrequency power of 250 W and at 13.56 MHz in an atmosphere of $Ar/O_2/N_2$ \cite{farhaoui2016reactive}.  The gas flows $F_{Ar}, F_{O2}$ and $F_{N2}$ are controlled using mass flowmeters. In the present work, the flows of $Ar$ and $N_2$ are put at a fixed value, while the flow of $O_2$ was made to vary periodically. More precisely, we use a 
a rectangular pulsed flow rate from 0 during the off-time $T_{off}$, to a maximum flow rate $F_{max O2}$ during the on-time $T_{on}$. Depending on pulsed parameters ($T_{on}$, $T_{off}$ and $F_{max O2}$), time-averaged $O_2$ flow rate is tuned during target sputtering which gives us a precise control on the composition of the deposited film, e.g. x, y and z in $Si_xO_yN_z$ material and so on their refractive index.

Two samples have been fabricated, corresponding to structures with a relatively low number of layers. The results presented above show that the physical mechanism which explain the efficiency of the ARC can actually occurs for as few as 6 layers, with only one period of the photonic crystal. The second sample corresponds to two periods of the Bragg mirror and thus 8 layers. These two structures are potentially cheap to manufacture and thus are more likely to be deployed some day than coatings with a thicker photonic crystal.

We measure the reflectance of the fabricated structures at room temperature. A halogen lamp is used as a visible light source. The collimated light is directed on the sample, then the reflected signal is focused on the monochromator slits entrance. A 32 cm focal monochromator coupled with a CCD camera is used as the detection.

\begin{figure}
    \centering
    \includegraphics[width=\linewidth]{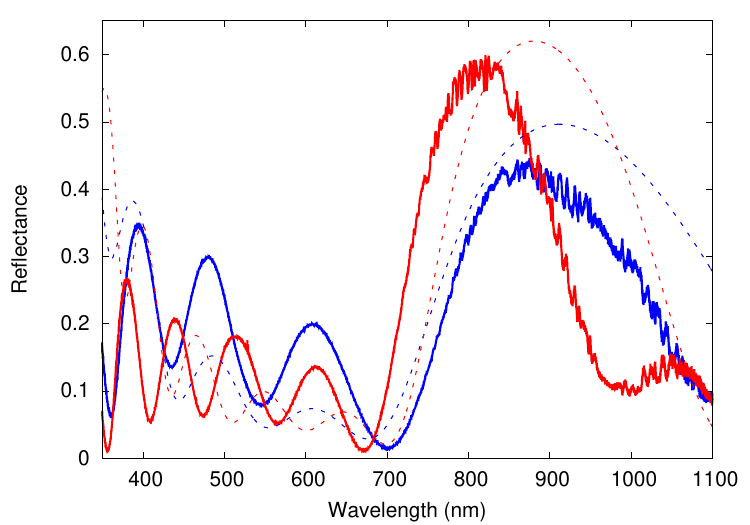}
    \caption{Experimental results of the reflectance measurements obtained for design with six layers (continuous blue line) and eight layers (continuous red line) and their theoretical expectation (discontinuous lines).}
     \label{fig:exp} 
\end{figure}

The measured spectra obtained for the two samples shown Fig. \ref{fig:exp} are in relatively good agreement with theoretical calculations, showing both a part of the spectrum presenting a low reflectance and a higher reflectance in the gap of the photonic crystal. We have little doubt that imperfections in the fabrication process, which is far from being optimized yet, are responsible for the discrepancies. At least, these results show how easily the ARC we have designed can be fabricated with easily deployable techniques.

\section{Conclusion} 

We shown that evolutionary optimization of anti-reflective coatings for photovoltaic devices systematically yields photonic crystals with adapted terminations. While such structures may not be perfectly optimal, we have established that they emerge for a very broad set of conditions. Such solutions can be called robust. While such an approach reminds of ripple reduction strategies which have been known for a very long time\cite{macleod2}, our work suggest that this class of structure should be the subject of much more attention. We insist that devices including photonic crystals whose working point is outside of the forbidden band, like the ARC studied here, are finally not that common\cite{centeno2005graded}. 

Light propagates in photonic crystals under the form of Bloch modes. Since their invention, Bragg mirrors have been thought as cavities for such a mode, and the ripples outside of the photonic bandgap as a signature of slab resonances. We have proposed a methodology to retrieve the internal reflection coefficient of the Bloch mode -- quantities which can not be retrieved analytically. First, our analysis is in complete agreement with the picture of finite structures as cavities. Second, our approach has confirmed that the outer layers of our ARC reduce the impedance mismatch between the outside media and the photonic crystal. The periodic part of the structures can thus be thought of as a buffer for light. 

This strategy is robust, in the sense that only the outer layers have to change when either the outer media or the index contrast change, explaining why such structures emerge so easily as solutions. Here, the physical analysis strengthens the confidence we have in the optimization process. 
A physical analysis is a powerful tool allowing to easily guess that a structure is probably sub-optimal or to perceive more precisely why it emerges\cite{barry2018evolutionary} and even to guide an optimization\cite{moreau07a,moreau12a}. As a consequence, we think the result of an optimization should always be discussed physically, which is only seldom done\cite{gondarenko2006spontaneous,frellsen2016topology}. 

Finally, we have shown that these structures can be easily fabricated using a sputtering process which is increasingly used in the industry. We underline that the bandgap of the photonic crystal present in our designs has the potential to reduce the heating of a potential solar cell, as it would reflect a part of the infra-red spectrum which is not always able to generate electron-hole pairs. This could be a real advantage for some photovoltaic devices when performances are more important than costs, like for concentration photovoltaics for instance\cite{xing2015review}.

We underline that we have been able to physically understand the structures produced by optimization because they are regular. In addition to being often a sign that the structure is close to optimality\cite{gondarenko2006spontaneous,barry2018evolutionary,tikhonravov2008estimation}, regularity generally allows for a physical analysis and thus, for the results of the optimization to be used as a true source of inspiration. As very few algorithms have in the past yield regular structures, the present work is one of the first example of such an {\em a posteriori} analysis. This is welcome because bio-inspiration does not always offer relevant solutions to technological problems, particularly in photonics. Being able to generate photonic structures which are technologically relevant through an evolution-like optimization process may be more promising.

\section{Acknowledgments}
The authors would like to thank Th\'eo Charreyron for his help. Computations have been performed on the supercomputer facilities of the M\'esocentre Clermont Auvergne University. Antoine Moreau is an Academy CAP 20-25 chair holder. He acknowledges the support received from the Agence Nationale de la Recherche of the French government through the program “Investissements d’Avenir” (16-IDEX-0001 CAP 20-25).

\bibliography{apssamp}

\end{document}



\title{SUPPLEMENTARY INFORMATION \\ Analysis and fabrication of antireflective coating for photovoltaics based on a photonic-crystal concept and generated by evolutionary optimization}


\author{Pauline Bennet}
\affiliation{Université Clermont Auvergne, CNRS, SIGMA Clermont, Institut Pascal,F-63000 Clermont-Ferrand, France}
\author{Perrine Juillet}
\affiliation{Université Clermont Auvergne, CNRS, SIGMA Clermont, ICCF, F-63000 CLERMONT-FERRAND, FRANCE}
\author{Sara Ibrahim}
\affiliation{Université Clermont Auvergne, CNRS, SIGMA Clermont, Institut Pascal,F-63000 Clermont-Ferrand, France}
\author{Vincent Berthier}
\affiliation{TAO, Inria, LRI, Université Paris Sud CNRS UMR 6823, Orsay Cedex, France}
\author{Mamadou Aliou Barry}
\author{François Reveret}
\affiliation{Université Clermont Auvergne, CNRS, SIGMA Clermont, Institut Pascal,F-63000 Clermont-Ferrand, France}

\author{Angélique Bousquet}
\affiliation{Université Clermont Auvergne, CNRS, SIGMA Clermont, ICCF, F-63000 CLERMONT-FERRAND, FRANCE}

\author{Olivier Teytaud}
\affiliation{Facebook AI Research, 6 rue Menars, 75000 Paris, France}
\author{Emmanuel Centeno}
\author{Antoine Moreau}%
\affiliation{Université Clermont Auvergne, CNRS, SIGMA Clermont, Institut Pascal,F-63000 Clermont-Ferrand, France}


\date{\today}

\maketitle
\renewcommand{\thefigure}{S\arabic{figure}}

\section{Emergence of the design}

\subsection{Budget choice}

\begin{figure}
    \centering
    \includegraphics[width=0.5\linewidth]{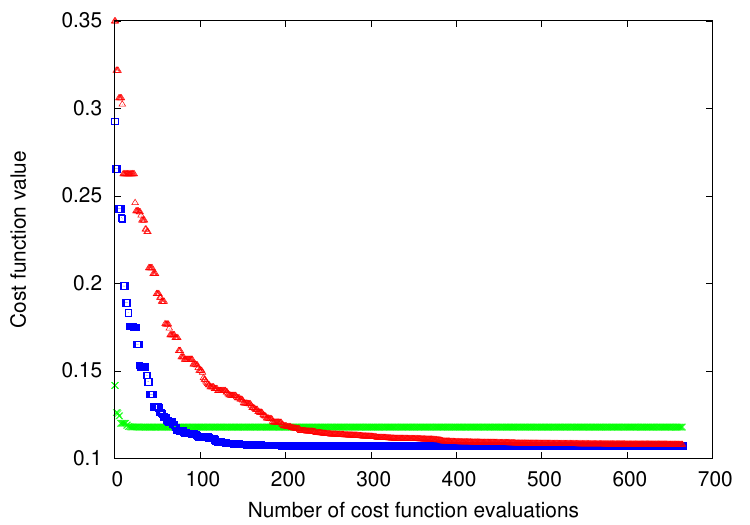} 
    \caption{Convergence curves obtained with DE and 2$\times$10$^4$ evaluations of the cost function for 3 situations, one with 2 layers (green), one with 26 layers (blue) and the third with 50 layers (red) for the ARC problem on silicon. This shows that this is a large enough budget for the algorithm to converge in each case. However, this does not mean that the optimum is reached, which is why we perform many runs for each situation.} 
    \label{fig:convergence_Sia_ref}
\end{figure}

Different algorithms can be compared only for the same number of evaluations of the cost function. Here we choose to allow 2$\times$10$^4$ evaluations for each run. This is generally enough for the DE to have converged towards a solution, as shown Fig.
\ref{fig:convergence_Sia_ref}.

\subsection{Broad emergence conditions}

We test several situations to study whether the same kind of structure consistently emerges, and when it does not because some constraints hinder the algorithm to converge towards it, whether the obtained solutions are better or not.
\begin{itemize}
    \item \textbf{Odd number of layers.} When forced to add another layer, the algorithm systematically puts this layer at the bottom of the structure, with a thickness as small as possible (30 nm being the minimum authorized). The cost function is  larger than without this layer as long as the optimization can be considered reliable. An even number of layer is thus clearly a better choice.
    \item \textbf{Beginning with the higher refractive index.} When the structure begins with the higher index on top and ends with the lower index at the bottom, while the optimization seems reliable, the cost function associated with the best solutions are typically 0.05 higher - a difference which can regularly be found between a random and an optimized structure. The solutions produced are in that case not periodic. This definitely shows that the solution which is physically the most sound (lower index in contact with air, higher index in contact with silicon) is actually the best.
    \item \textbf{Leave the refractive index vary.} We have left the algorithms change the refractive index of the layers. While the algorithm understandably struggles for a lower number of layers the reference solution is found whenever the algorithm is still reliable. This result is important because it shows that the solution emerges whenever the available refractive index are constrained, and that it is not produced by the arbitrary choice for the refractive index\cite{barry2018evolutionary}. And quite unsurprisingly, it shows that taking the highest possible index contrast is the best choice anyway, for this index contrast at least.
    \item \textbf{Increasing the index contrast.} We have imposed a greater contrast on the refractive indexes, choosing values typically considered for the design of optical filters\cite{tik1996}. The results are clearly better than for a lower index contrast, by almost one order of magnitude - however the structures produced consist in a photonic crystal with adapted outer layers too.
    \item \textbf{Changing the substrate.} We test all the cases above for three other different substrates, (i) an infinite crystalline silicon (ii) an infinite silver layer below a thin (89 nm) thickness of amorphous silicon and (iii) an infinite silver layer below a thick (1000 nm) thickness of amorphous silicon. In all cases, the results converge towards a photonic crystal with one, two or eventually three adapted layers at the extremities.
    \item \textbf{Optimization of the extremities of the design only.} We impose a central photonic crystal in the structure and try to optimize only the four (respectively six) external layers. The optimization leads to a thicker photonic crystal with one (respectively two) additional periods in the photonic crystal, and only two adapted layers in each cases.
\end{itemize}

All our optimizations thus points towards a photonic crystal with adapted external layers as the best solution, except in one case: when the index contrast is large and the refractive index of each layer is left to vary (see below for such a case). 

\subsection{Analysis of the best structure produced in standard conditions}

\begin{figure}
\includegraphics[width=0.5\linewidth]{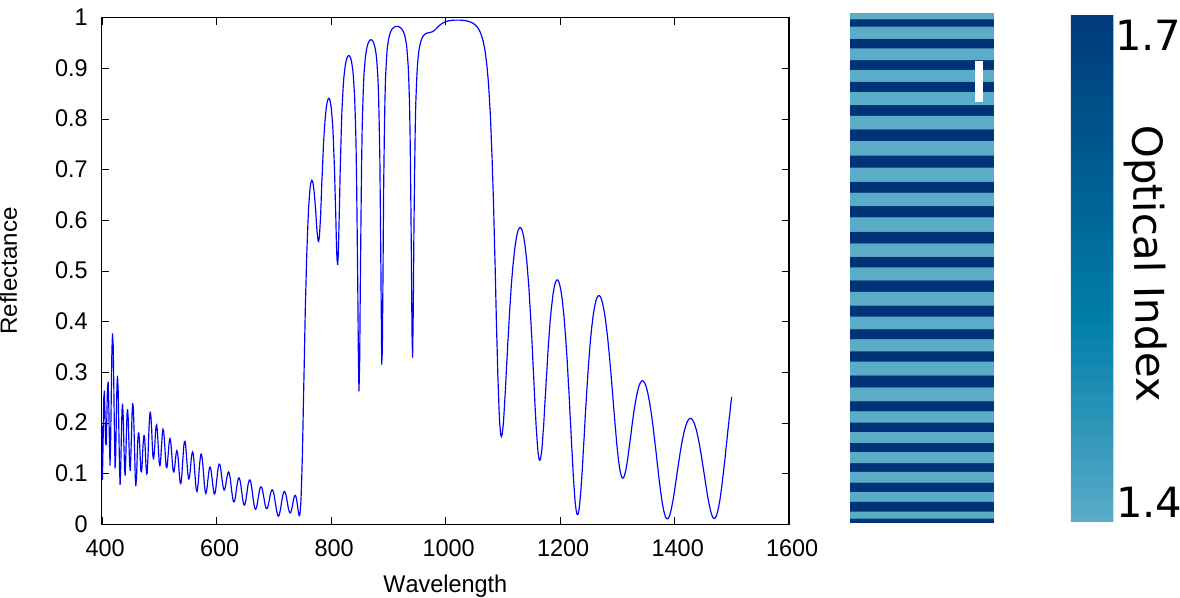}
    \caption{Reflectance spectrum and representation of the structure with 46 layers. The white line represents 500 nm.}
    \label{fig:spectrum_structure_46_ref_Sia}
\end{figure}

In all our runs, the very best result in standard conditions is obtained for 46 layers. The structure is a slightly more disordered photonic crystal, thus presenting some defects, whose signature can be seen in the forbidden band as dips in the reflectance as shown Fig. \ref{fig:spectrum_structure_46_ref_Sia}. Anyway, even though the optimization results are not perfect, they point towards a photonic crystal based solution too. This is clear when the structure is considered or when analyzing the spectrum.

\subsection{Leave the refractive index vary for a large index range}

\begin{figure}
\includegraphics[width=0.5\linewidth]{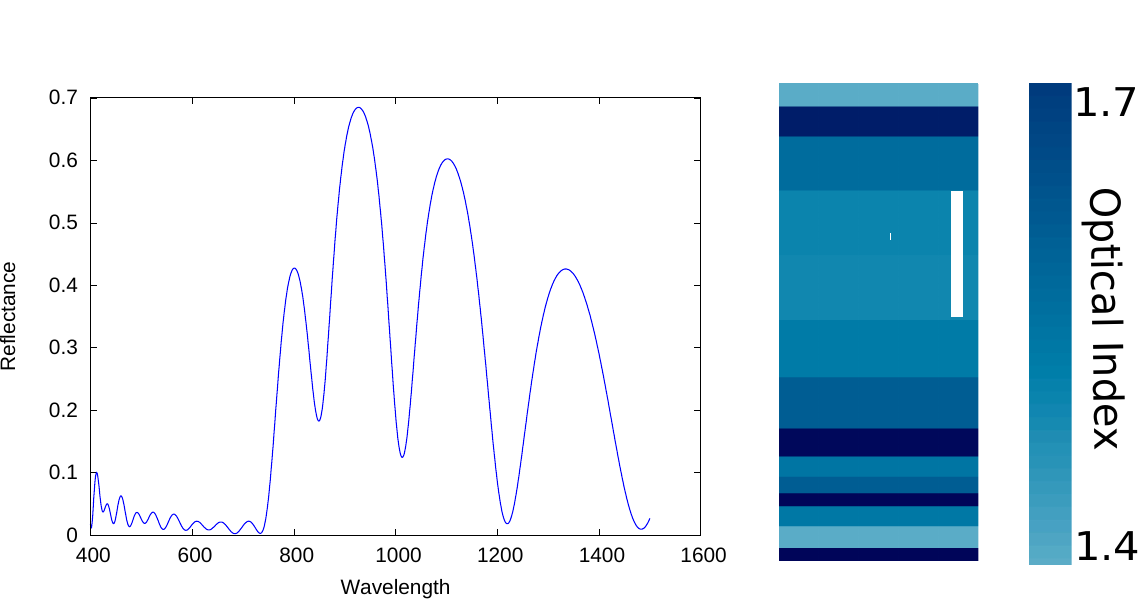}
    \caption{Reflectance spectrum and representation of the structure with 14 layers, when the refractive index can vary on a large contrast. The white line represents 500 nm.}
    \label{fig:spectrum_structure_14_GC_nvar}
\end{figure}

\begin{figure}
\includegraphics[width=0.5\linewidth]{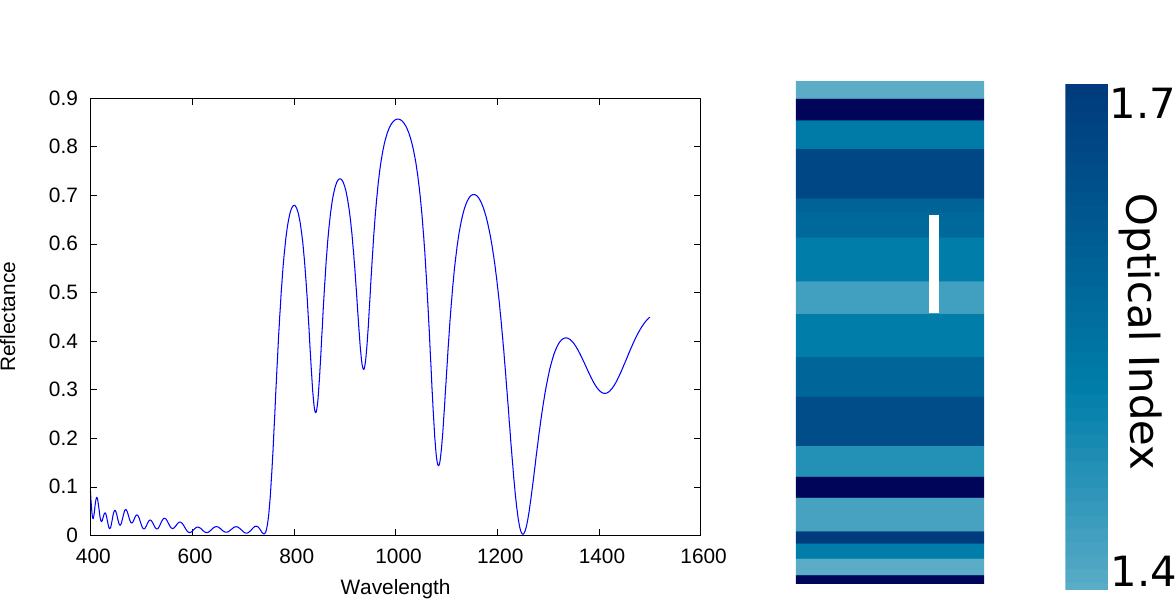}
    \caption{Reflectance spectrum and representation of the structure with 18 layers, when the refractive index can vary on a large contrast. The white line represents 500 nm.}
    \label{fig:spectrum_structure_18_GC_nvar}
\end{figure}

The photonic crystal (PhC) based ARC is not the best possible solution whatever the conditions. We have run optimizations for a large refractive index range, but we have let the algorithm change the refractive index of all layers. Our data indicate that such a problem is a complicated one for the algorithm, which most often does not yield perfectly reliable solutions (no plateau for the values of the cost function, or a higher cost function than for the PhC based structure). However two solutions particularly stand out. They are slightly better than the PhC based ones but do not present layers alternating from the highest possible index to the lowest (see Fig. \ref{fig:spectrum_structure_14_GC_nvar} and \ref{fig:spectrum_structure_18_GC_nvar}). In this case, the results differ completely from the PhC based solution. However, they present very regular and interesting structures which thus seem at first to obey a completely different logic. For instance, they challenge the widespread belief that the best strategy for any kind of optical filter, is to alternate between the highest and the lowest available index. The inner symmetrical structure that both these solutions present is surprising.

Furthermore, this result strengthen the idea that DE is able to produce any kind of regular solutions, periodic or not at all, in this case. The physical analysis of these structures is left for future work, but we underline that they too reflect a large part of the infra-red spectrum even if no photonic crystal is present.

\section{Physical analysis through a cavity model for the Bloch mode}

\subsection{Details on the Fabry-Perot model}

By modeling the PhC as a Fabry-Perot cavity, we can compute the reflectance of the whole structure with the formula 
\begin{equation}
    r(\lambda, h)=r_0(\lambda)+\frac{t_1(\lambda) t_2(\lambda) r_{Bs}(\lambda) e^{2i\gamma(\lambda) h}}{1-r_{Bi}(\lambda) r_{Bs}(\lambda) e^{2i\gamma(\lambda) h}},
\end{equation}
with $r_0$ the reflection of the incident plane wave on the first layer of the photonic crystal, $t_1$ and $t_2$ the transmission coefficients between free space and the Bloch mode, $r_{Bi}$ (resp. $r_{Bs}$) is the reflection coefficient of the Bloch mode on the outer layers (resp. the substrate), $h$ the thickness of the photonic crystal, and $\gamma$ the constant of propagation of the Bloch mode inside the PhC. 

The model can be applied even if we consider the adapted layers of the simplified design, which makes the whole structure thicker. In the case we keep defining the thickness of the cavity as the thickness of the photonic crystal core. Such a choice is arbitrary, but we underline that a different choice would only lead to a change of the phase of the internal reflection coefficients for the Bloch mode. The two adapted layers can be represented by a single scattering matrix, acting as some kind of peculiar interface. The addition of the adapted layers in our system thus leads to the same Fabry-Perot formula as for a finite Bragg mirror for the global reflection coefficient $r$. Only the values of the internal transmission or reflection coefficients are modified. Our physical analysis relies on the change brought to these very coefficients by the presence of the adapted layers.

\subsection{Retrieval of the internal reflection coefficients}

Not all the $c_i$ coefficients are however uniquely determined in the fitting process. We consider for instance all the structures beginning with two air-adapted layers. Physically, we could expect the coefficient $c_0$, corresponding to the reflection coefficient $r_0$ on the top of each structure to be the same in all these cases. However, the coefficient is uniquely defined only for the structure which is symmetric -- this one ending by two air-adapted layers. For the three other cases, $c_0$ oscillates between a curve specific to each structure and the coefficient computed for symmetric case. This suggests the coefficient computed for the symmetric case should be considered the right one. In the paper, however, we focus on the $c_2$ coefficient, which is always uniquely defined and allows to access the internal reflection coefficients of the Bloch mode itself, thus providing enough insight to understand the role of the outer layers.

\begin{figure}
    \centering
    \includegraphics[width=0.5\linewidth]{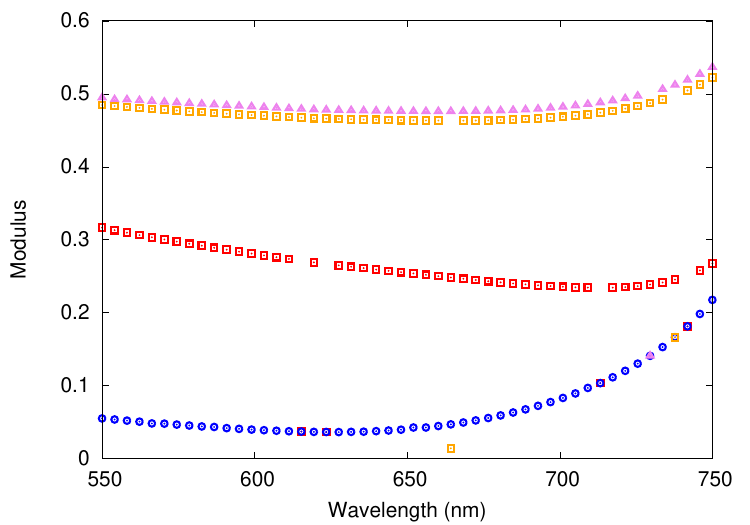}
    \caption{Modulus of $r_0$ parameter with air-adaptive layers.
    Pink triangles : incident plane wave reflection given by structure d.
    Orange squares: incident plane wave reflection given by structure e.
    Red squares : incident plane wave reflection given by structure a.
    Blue circles : incident plane wave reflection given by structure b. 
    }
    \label{fig:r0_blue}
\end{figure}

\begin{figure}
    \centering
    \includegraphics[width=0.5\linewidth]{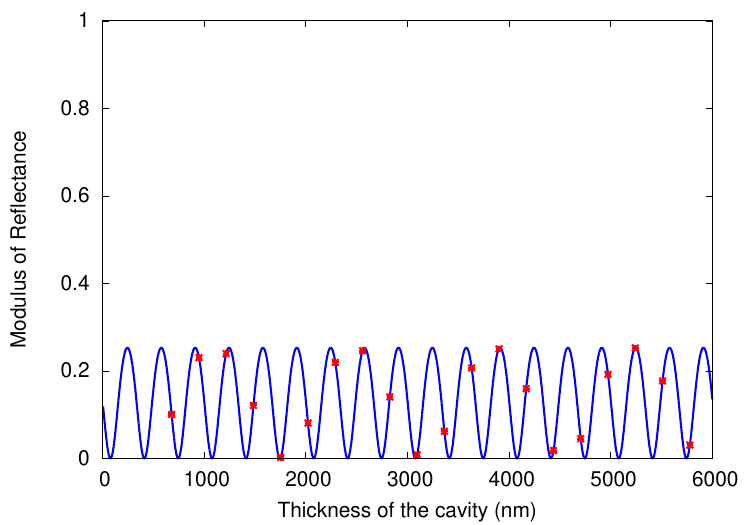}
    \caption{Representation of all the designs studied. Each curved arrow represents one Bloch mode reflection coefficient from photonic crystal toward an external media. The design of interest is this one represented on (a). The other designs allow the study of the behavior of the Bloch mode in function of the external media and to understand why the (a) design is the solution of the photovoltaic design optimization. A yellow layer represents a silicon substrate.}
    \label{fig:all_structures}
\end{figure}

However, even if $c_2$ is uniquely defined, it provides an immediate access to the internal reflection coefficients only in the case of a symmetric structure. In general, $c_2$ is the product $r_{Bi} r_{Bs}$ and thus the determination of both coefficients is not straightforward. In fact, determining these coefficients is possible and constitutes a coherent approach, provided the fitting process is applied to a variety of symmetric and asymmetric structures (see Fig.\ref{fig:all_structures}).

As explained above, the only way to access an internal reflection coefficient for the Bloch mode is in the case of a symmetric structure. It is given in that case by $\sqrt{c_2}$. Here, we will use symmetric structures to retrieve the internal reflection coefficient and check the result is coherent with the parameters retrieved for asymmetric structures. The different structures we have considered are all shown Fig. \ref{fig:all_structures}.

\begin{figure}
    \centering
    \includegraphics[width=0.5\linewidth]{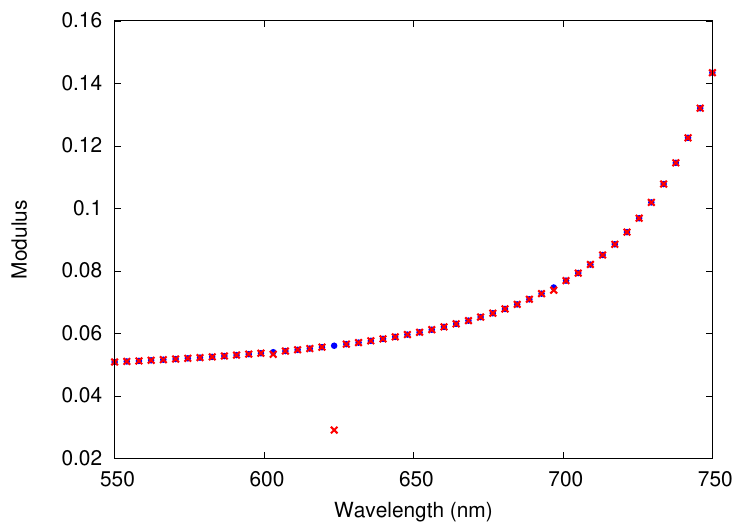}
    \caption{Comparison of the modulus of $c_2$ for asymmetric case (g) (red dots) with the product of the reflection coefficients for the Bloch modes retrieved using the symmetric cases (f) and (h) (blue dots). The point clearly outside of the curve corresponds to a failed optimization.}
    \label{fig:sanityFGHmod}
\end{figure}

In order to check the coherence of our approach, we begin with the simple Bragg mirror. A finite Bragg mirror can be terminated by a high or low index layer, on each side. This leaves three different combinations (f) (g) and (h) shown Fig. \ref{fig:all_structures}. The case (f) (respectively (h)) allows to retrieve the internal reflection coefficient when the outer layer has a low (respectively high) refractive index. When these two coefficients are multiplied by each other, we find exactly the value of the $c_2$ coefficient for structure (g) whatever the wavelength (see Fig. \ref{fig:sanityFGHmod}). This means that the reflection coefficients we have found for symmetric cases can actually be interpreted as the reflection coefficients of the Bloch modes and that modeling the finite Bragg mirror as a cavity makes perfectly sense.

In order to better understand the role of the adapted layers on top and at the bottom of the Bragg mirror in our reference structure (case (a)), we need to make sure we can determine the corresponding internal coefficients uniquely. 

\begin{figure}
    \centering
    \includegraphics[width=0.5\linewidth]{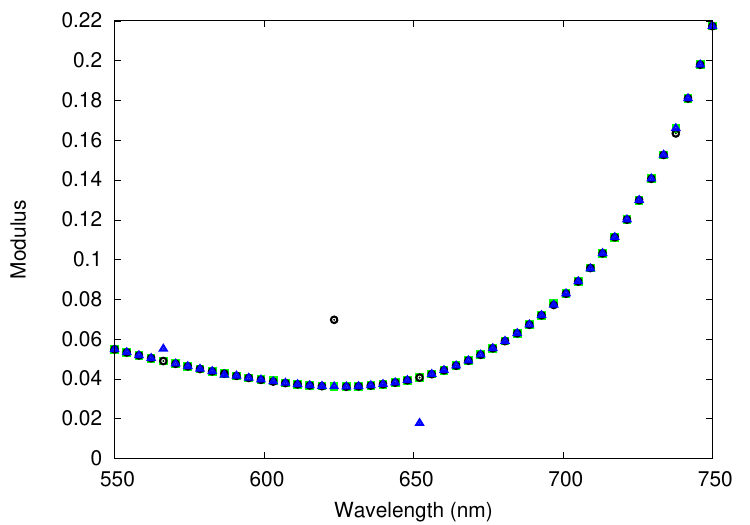}
    \caption{Modulus of the reflection for the Bloch mode on air-adaptive layers. Blue triangles : air-adaptive Bloch reflection given by structure (b). Black circles : air-adaptive Bloch reflection given by structure (m). Green squares : air-adaptive Bloch reflection given by structure (k). }
    \label{fig:blue_verif}
\end{figure}

We begin with the adapter layers on top of the structure. This coefficient can be determined by considering the symmetric structure (b). In order to check that this is a coherent approach we have considered the asymmetric structures (k) and (m) which present on one side adapted layers and an abrupt termination of the Bragg mirror on the other side. The internal reflection coefficients computed for the abrupt terminations of the Bragg mirror provide in these case another way to determine the reflection coefficient with the adapted layers. Fig. \ref{fig:blue_verif} shows these three different ways to retrieve the reflection coefficient of the Bloch mode provide the same result.

\begin{figure} 
    \centering
    \includegraphics[width=0.5\linewidth]{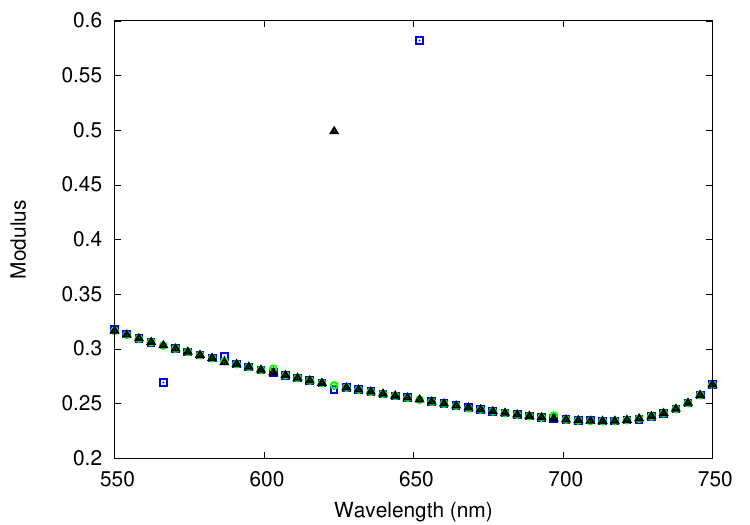}
    \caption{Modulus of the reflection for the Bloch mode on silicon-adaptive layers. Black triangles : silicon-adaptive Bloch reflection given by structure (l). Green circles : silicon-adaptive Bloch reflection given by structure (n). Blue squares : silicon-adaptive Bloch reflection given by structure (a).}
    \label{fig:red_verif}
\end{figure}

Retrieving the reflection coefficient on the silicon side with adapted layers can be done following the same methodology -- except that since silicon is an absorbing medium, it is not possible to imagine a symmetric structure which could be illuminated. However we can use asymmetric cases (l), (n) and (a) with abrupt or adapted terminations to retrieve this coefficient. Fig. \ref{fig:red_verif} shows that the three cases yield the same result.

More generally, we have check for all the structures shown Fig. \ref{fig:all_structures} that our approach produces perfectly coherent results, allowing us to access the reflection coefficient of the Bloch mode whatever the termination for the Bragg mirror considered (abrupt or adapted, with silicon or with air). 

\begin{figure}
    \centering
    \includegraphics[width=0.5\linewidth]{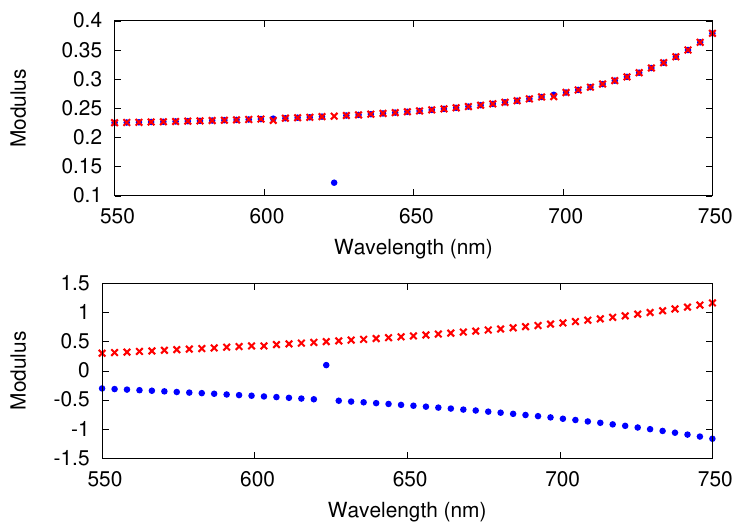}
    \caption{Modulus (top) and phases (bottom) comparison between the coefficient of reflection for the Bloch mode inside a symmetric Bragg mirror beginning with a low refractive index (blue points) and beginning with a high refractive index (red crosses).}
    \label{fig:impactFirstLayer}
\end{figure}

Finally, we would like to underline that the reflection coefficient of the Bloch mode in the case of an abrupt termination of the photonic crystal presents the same modulus, whether the outer layer has a lower or higher refractive index (Fig.\ref{fig:impactFirstLayer}). Only the phase changes: they are opposite from each other. We have not actual physical explanation, but we think that having access to these internal reflection coefficients could be useful to better describe the Bragg mirrors as a homogeneous medium.

\bibliography{supp}